\documentclass[aps,prl,twocolumn,superscriptaddress,amsfont,graphicx,nofootinbib,preprintnumbers]{revtex4-1}%
\usepackage{color,graphicx,epsfig}
\usepackage{ifpdf}
\usepackage{amsmath}
\usepackage{bm}
\usepackage{color}
\usepackage[english]{babel}
\usepackage{graphicx}%
\usepackage{amsfonts}%
\usepackage{amssymb}
\usepackage{braket}
\usepackage{hyperref}
\usepackage{enumerate}
\usepackage{comment}

\definecolor{nicered}{rgb}{0.7,0.1,0.1}
\definecolor{nicegreen}{rgb}{0.1,0.5,0.1}
\hypersetup{colorlinks,citecolor= nicegreen,linkcolor= nicered}

\usepackage{graphicx}
\usepackage{dcolumn}
\usepackage{bm}
\usepackage{slashed}

\begin{document}

\title{Top quark as a probe of heavy Majorana neutrino at the LHC and future collider}

\author{Ning Liu}
\email{liuning@njnu.edu.cn}
\affiliation{Department of Physics and Institute of Theoretical Physics, Nanjing Normal University, Nanjing, 210023, China}

\author{Zong-guo Si}
\email{zgsi@sdu.edu.cn}
\affiliation{School of Physics, Shandong University, Jinan, Shandong 250100, China}

\author{Lei Wu}
\email{leiwu@njnu.edu.cn}
\affiliation{Department of Physics and Institute of Theoretical Physics, Nanjing Normal University, Nanjing, 210023, China}

\author{Hang Zhou}
\email{zhouhang@njnu.edu.cn}
\affiliation{Department of Physics and Institute of Theoretical Physics, Nanjing Normal University, Nanjing, 210023, China}

\author{Bin Zhu}
\email{zhubin@mail.nankai.edu.cn}
\affiliation{School of Physics, Yantai University, Yantai 264005, China}
\affiliation{Department of Physics, Chung-Ang University, Seoul 06974, Korea}

\date{\today}

\begin{abstract}
Right-handed (RH) Majorana neutrinos play a crucial role in understanding the origin of neutrino mass, the nature of dark matter and the mechanism of matter-antimatter asymmetry. In this work, we investigate the observability of heavy Majorana neutrino through the top quark neutrinoless double beta decay process $t \to b \ell^+ \ell^+ j j$ at hadron colliders. By performing detector level simulation, we demonstrate that our method can give stronger limits on the light-heavy neutrino mixing parameters $|V_{eN, \mu N}|$ in the mass range of 15 GeV $< m_N <$ 80 GeV than other existing collider bounds.
\end{abstract}
\pacs{Valid PACS appear here}
\maketitle


\section{Introduction}
The discovery of neutrino oscillations in solar, atmospheric, reactor and accelerator experimental data indicates that neutrinos are massive and mixed~\cite{Tanabashi:2018oca}. This stands for a robust evidence for new physics beyond the Standard Model (SM).  The seesaw mechanism provides a natural framework for generating a small Majorana neutrino mass~\cite{minkowski1977,yanagida1979,ms1980,sv1980,Weinberg:1979sa,mw1980,cl1980,lsw1981,ms1981,flhj1989,ma1998}. There are many models that could incorporate the Majorana mass, such as SO(10) Supersymmetric (SUSY) grand unification~\cite{Harvey:1981hk,Dimopoulos:1991yz,Hall:1995eq} and other grand unified theories (GUT)~\cite{Dorsner:2005fq,Dorsner:2005ii,Dorsner:2006hw}, which naturally links the tiny neutrino masses with new physics at the GUT-scale. Thus, the typical scale for heavy Majorana neutrino mass $M_N$ in GUT is of order the GUT-scale.

Among various seesaw models, the Type-I seesaw extension of the SM is the simplest version, in which three singlet RH neutrinos ($N_{R\alpha}$) are included~\cite{minkowski1977,yanagida1979,ms1980,sv1980}. The full mass term of neutrinos can be written as,
\begin{align}
-{\cal L}_M = y_{i \alpha} \bar{L}_{Li} \widetilde{\Phi} N_{R\alpha} + \frac{1}{2}(M_N)_{\alpha \beta} \bar{N}_{R\alpha}^C N_{R\beta} +{\rm H.c.} \; ,
\label{lagM}
\end{align}
where $i=1,2,3$ is the generation index and $\alpha=e,\mu,\tau$ is the RH neutrino flavor index. After the electroweak symmetry breaking, the neutrino mass matrix in the flavor basis $\{\nu^C_{Li},N_{R\alpha}\}$ is given by,
\begin{align}
\label{eq:seesaw}
	{\cal M}_\nu \ = \ 		\begin{pmatrix}
			\mathbf{0}   & M_D \\
				M_D^{\sf T} & M_N
		\end{pmatrix} \; .
\end{align}
The diagonalization of Eq.~\ref{eq:seesaw} leads to the mixing between light and heavy neutrinos. Then, the light-flavor neutrinos $\nu_{i L}$ can be expressed as a combination of light and heavy mass eigenstates,
\begin{eqnarray}
\nu_{i L}=\sum^{3}_{m=1}U_{im}\nu_{m L}+\sum^{6}_{m=4}V_{im}N^{c}_{m L},
\end{eqnarray}
where the mixing matrices $U$ and $V$ satisfy the unitary condition $UU^{\dag}+VV^{\dag}=I$. If $M_N$ is of order GUT-scale, the light-heavy neutrino mixing parameters $V^2_{i N} \sim M_D / M_N$ are usually too tiny to produce sizable effects in various physical processes.  However, if $M_N$ is allowed to be much lower, then the induced effects of such heavy Majorana neutrinos can be searched for in some existing experiments.  Therefore, it is crucial to explore the possibilities of implementing seesaw mechanism at low energies from both theoretical and experimental researches.

Recently,  many low-scale Type-I seesaw scenarios~(see examples,~\cite{Asaka:2005an,Asaka:2005pn,Asaka:2006nq,Asaka:2006ek,Xing:2009in,Adhikari:2010yt,Ibarra:2010xw,Boucenna:2014zba,Zhou:2017lrt,Gu:2018kmv}) have been proposed, in which $M_N$ may be accessible in foreseeable experiments~\cite{deGouvea:2006gz,deGouvea:2007hks,Atre:2009rg,Kersten:2007vk,Bajc:2007zf,He:2009ua,Han:2006ip,Ibarra:2011xn,Dev:2013wba,Deppisch:2015qwa,Das:2018hph}. This provides an unique opportunity to test the link between the origin of neutrino mass and the observed matter-antimatter asymmetry via the leptogenesis at colliders (for a review, see e.g.~\cite{Davidson:2008bu}). As can be seen from Eq.~\ref{lagM}, generating Majorana neutrino masses will at the same time lead to violation of lepton number by $\Delta L = 2$. The lepton-number-violation (LNV) processes can serve as smoking gun to test the mechanism of neutrino mass generation. One promising probe for Majorana neutrino is through the neutrinoless double beta decay ($0\nu\beta\beta$)~\cite{Furry:1939qr,Doi:1985dx}, which gives the most stringent bound on the mixing parameter $V_{eN}$~\cite{Agostini:2018tnm}. By fitting the electroweak precision observables, the mixing parameters $V_{\mu N}$ and $V_{\tau N}$~\cite{delAguila:2008pw} can be constrained tightly. In addition, a variety of low energy processes with $\Delta L=2$ can also be used to probe Majorana neutrinos~(see example,~\cite{Atre:2009rg}), such as decays of $\tau$, mesons\,($\pi$, $K$, etc.)~\cite{Ng:1978ij,Abad:1984gh,Dib:2000wm,Ali:2001gsa} and hyperons\,($\Sigma$, $\Xi$, etc.)\,\cite{Littenberg:1991rd,Barbero:2002wm}. However, these low-energy observables can not completely prove the Majorana nature of heavy neutrinos because models with pseudo-Dirac heavy neutrinos can also produce the same effects~(see example,~\cite{Malinsky:2009df}). Therefore, it is essential to perform an independent direct search for heavy neutrinos at colliders.

Heavy neutrinos that have masses below TeV scale in the low-scale Type-I seesaw can be directly produced at colliders (for a review, see e.g.~\cite{Deppisch:2015qwa}). By searching for the process $e^+e^-\to N(\to \ell W,~\nu_{L_i} Z,~\nu_{L_i} H) \nu_{L_i}$, the LEP experiment puts a 95\% C.L. upper limit on the mixing parameter $|V_{eN,\mu N}|^2<{\cal O}(10^{-5})$ in a heavy neutrino mass range between 80 and 205 GeV~\cite{Abreu:1996pa}. Recently, the CMS collaboration has performed searches for heavy Majorana neutrinos through the Drell-Yan process $q\bar{q} \to W^* \to N\ell$ and photon initial process $\gamma q \to W^* q' \to N \ell q'$ in trilepton and same-sign dilepton final states, which give the current stringent limits on $|V_{eN,\mu N}|^2$ from ${\cal O}(10^{-5})$ to unity in the masses of heavy neutrinos between 20 GeV and 1600 GeV~\cite{Sirunyan:2018mtv,Sirunyan:2018xiv}. 

As a top-rich environment, the Large Hadron Collider (LHC) can produce copious top quark events and may give a good opportunity to test the low-scale seesaw models~\cite{BarShalom:2006bv,Si:2008jd,Alcaide:2019pnf}. In this work, we will demonstrate that the top quark neutrinoless double beta decay process $t \to bjj\ell^{+}\ell^{+}$ (see Fig.~\ref{fig:topdecaybr}) provides a new way to search for the GeV scale-electroweak scale heavy Majorana neutrino, which will give the signature of the same-sign dilepton plus multi-jet through $t\bar{t}$ production at the LHC. In the following calculations, we will show that our strategy has a better sensitivity of probing the light-heavy neutrino mixing parameter $V_{\mu N}$ than other existing methods when 15 GeV $\lesssim m_N \lesssim$ 80 GeV.

\section{Search for \texorpdfstring{$0\nu\beta\beta$}{Lg} decay of top quark}

\begin{figure}
\centering
\includegraphics[width=7cm,height=6cm]{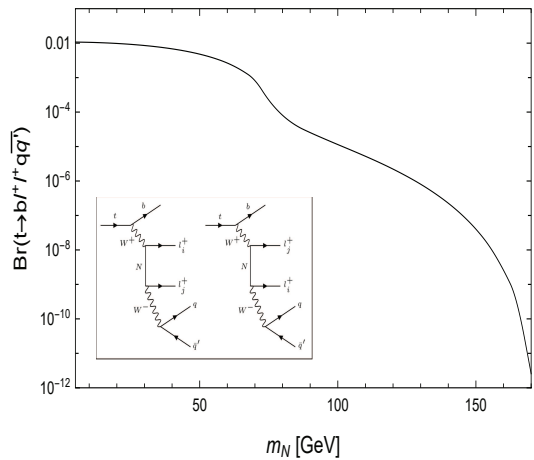}
\caption{Normalized branching ratio of top rare decay $t \to b l^{+}l^{+}q\bar{q'}$.}
\label{fig:topdecaybr}
\end{figure}
As a phenomenological study, we will parameterize the low-scale Type-I seesaw as a single RH Majorana neutrino mass scale $M_N$ and a single flavor light-heavy neutrino mixing $V_{i N}$. Such a framework allows us to remain agnostic of the detailed UV-physics, yet still capture the feature of low-scale Type-I seesaw.
The effective interactions between Majorana neutrinos and charged leptons in the mass eigenstates is given by,
\begin{align}
\mathcal{L}={}&-\frac{g}{2\sqrt{2}}V_{ij}W^{+}_{\mu}l_{i}\gamma^{\mu}(1-\gamma_{5})N^{c}_{j}+\textrm{H.c.}
\label{L_NLW}
\end{align}
In our study, we assume that the mixing effects in other flavors $\ell'\neq \ell$ are sub-dominant. This will enables us to derive generic bounds on the mixing parameter, which can be translated or scaled appropriately in the context of particular neutrino mass models. Besides, we will focus on the semileptonic decay of the $W$ boson, because it is impossible to determine whether the leptonic channel is induced by the Majorana neutrino.

In Fig.~\ref{fig:topdecaybr}, we present the dependence of the normalized branching ratio of the top rare decay channel $t \to b\,l^{+}l^{+}q\bar{q'}$ on the Majorana neutrino mass $m_{N}$, where the effective mixing $C_{ij}\equiv |V_{iN}V_{jN}|^{2}/ \sum_{k=e,\mu,\tau}|V_{kN}|^{2}$. From Fig.~\ref{fig:topdecaybr}, we can find that the normalized branching ratio can be as large as $10^{-4}$ to $10^{-2}$ when 15 GeV $<m_N < m_W$. With the increase of $m_N$, the branching ratio will decrease rapidly due to the suppression of phase space. As such, we will focus on the kinematical region of $m_N < m_W$ in our study.

In order to understand the mass generation mechanism and flavor structure of neutrino sector, it is essential to measure the mixing parameter and mass of each kind of neutrino at the LHC. It should be noted that the limit on the mixing parameter $|V_{\tau N}|^2$ strongly depends on the identification of same-sign di-tau, which is quite difficult and has low efficiency in the realistic collider simulation at the LHC. Therefore, we estimate that the current LHC sensitivity for the mixing parameter $V_{\ell N}$ ($\ell=e,~\mu$) can be improved significantly for the heavy neutrino mass range of interest, i.e. 15 GeV $<m_N < m_W$. Then, we will carry out the Monte Carlo simulation of the following signature,
\begin{equation}
pp \to t\bar{t} \to 2b+\ell^+ \ell^+ +4j
\end{equation}
where $t \to b \ell^+ \ell^+ j j$ and $\bar{t} \to b j j$. The contribution of the process $pp \to t\bar{t} \to 2b+\ell^- \ell^- +4j$ is also included. Since there are two same sign leptons (2SSLs) plus multi-jets in our signal, the main SM backgrounds include:
\begin{itemize}
\item multiple prompt leptons: they mainly come from events with two vector bosons, such as $W^\pm W^\pm$+jets and $t\bar{t}W^\pm$. Besides, the processes of $WZ$+jets and $ZZ$+jets can lead to 2SSLs, if one or more of the leptons fail the reconstruction or selection criteria.
\item misidentified leptons: The fake leptons can be misidentified hadrons that are from heavy-flavor jets. These fake leptons are generally less isolated than a prompt lepton from a $W/Z$ boson decay. The main contribution arises from $t\bar{t}$ events.
\item sign mismeasurement: The events that have two opposite-sign leptons with jets could contaminate our signal due to the mismeasured sign of leptons. But the mismeasurement rate of the sign of an electron or muon is usually small, which will be not considered in this analysis.
\end{itemize}
Besides, it should be mentioned that the leptons arising from $b$ and $c$ decay can also mimic our signals because of the large heavy flavor production cross section. However, these leptons are usually too soft to satisfy our criterion of two prompt same sign leptons with $P_T>10 GeV$. According to the CMS analysis, their contributions are expected to be negligible~\cite{Sirunyan:2018mtv}.

We generate the parton-level signal and background events by using \textsf{MadGraph5\_aMC@NLO}~\cite{Alwall:2014hca}. Within the framework of \textsf{CheckMATE2}~\cite{Dercks:2016npn}, we then implement parton showering and hadronization by \textsf{Pythia-8.2}\,\cite{pythia}, while the detector effects are simulated by tuned \textsf{Delphes3}~\cite{delphes}. Jets-clustering is done by \textsf{FastJet}~\cite{fastjet} with the anti-$k_t$ algorithm~\cite{anti-kt}. In the simulation we assume a $70\%$ b-tagging efficiency. To include the higher order QCD corrections, we normalize the leading order cross sections of $t\bar{t}$ and $t\bar{t}W^\pm$ to their NNLO and NLO values, respectively~\cite{Czakon:2011xx,Frixione:2015zaa}.

\begin{figure}[ht]
  \includegraphics[width=4.25cm,height=4.5cm]{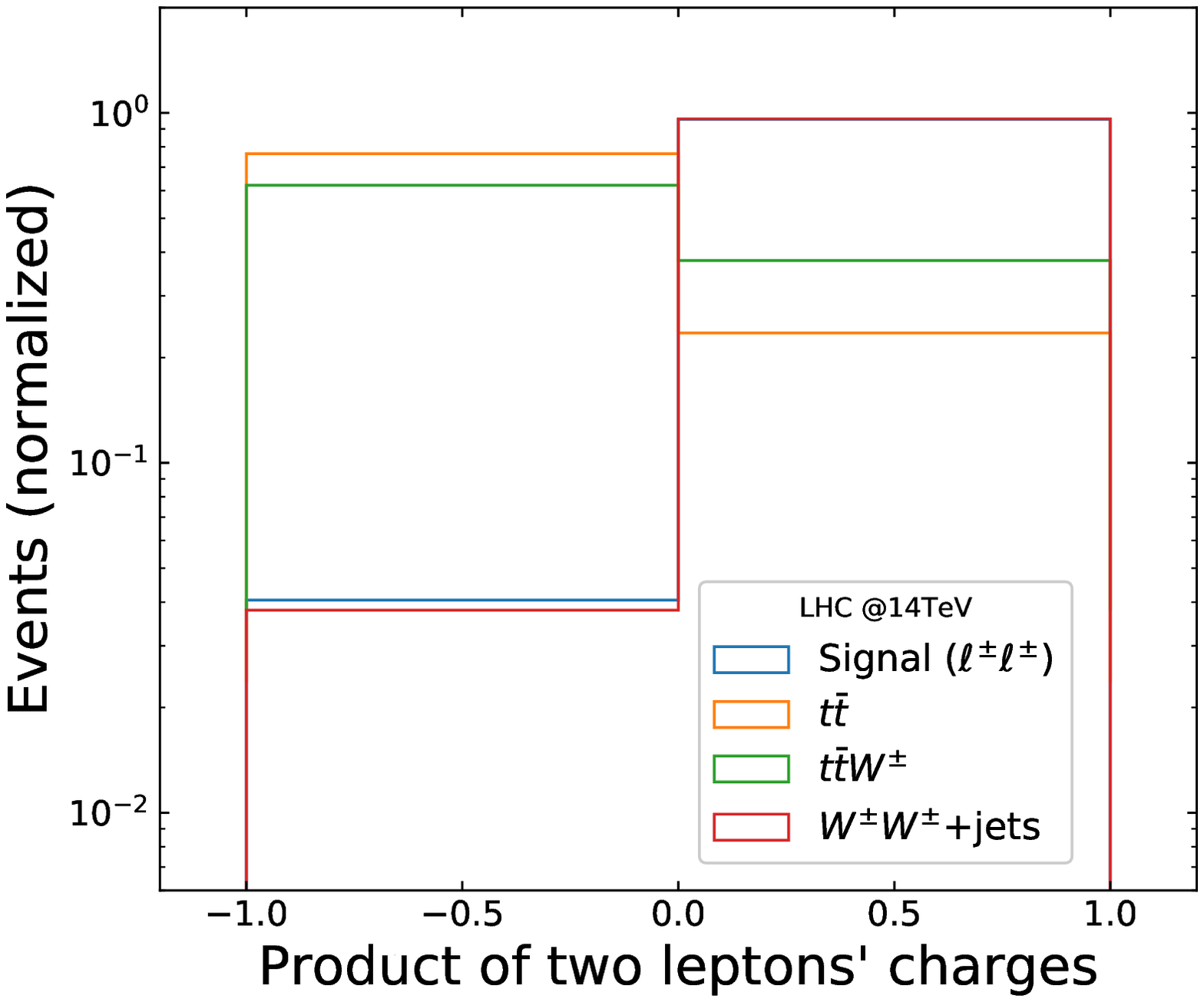}
  \includegraphics[width=4.25cm,height=4.5cm]{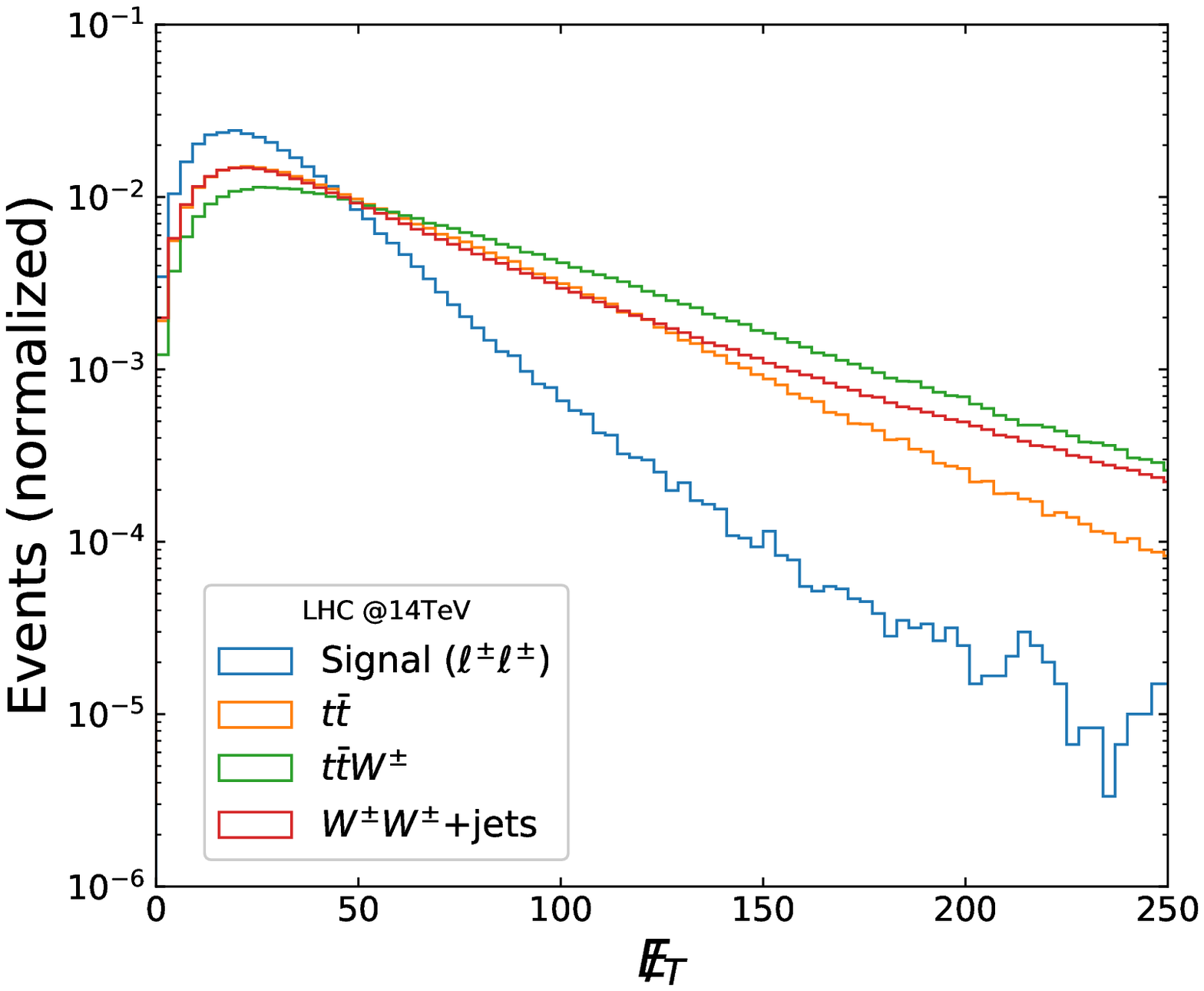}
  \includegraphics[width=4.25cm,height=4.5cm]{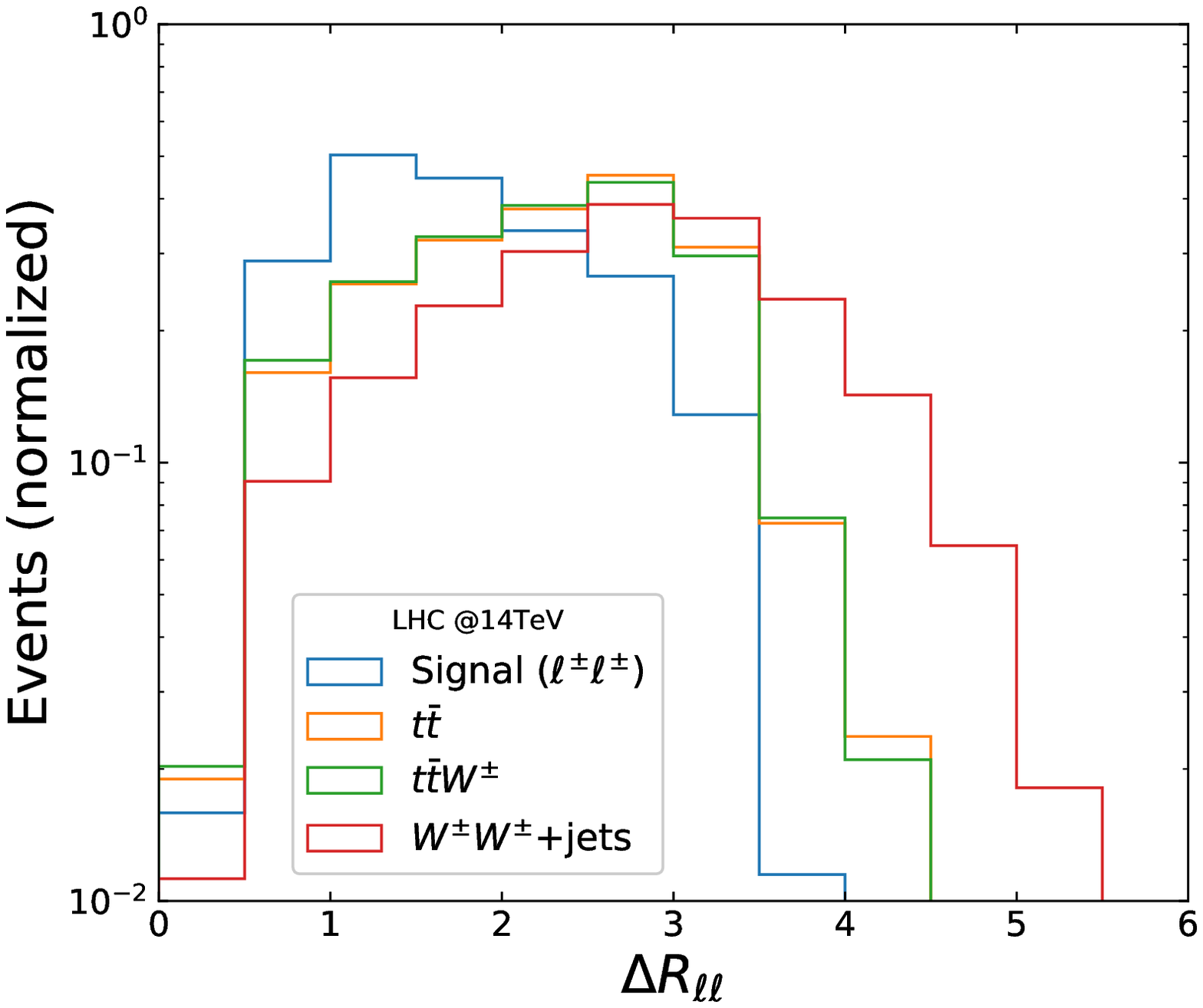}
  \includegraphics[width=4.25cm,height=4.5cm]{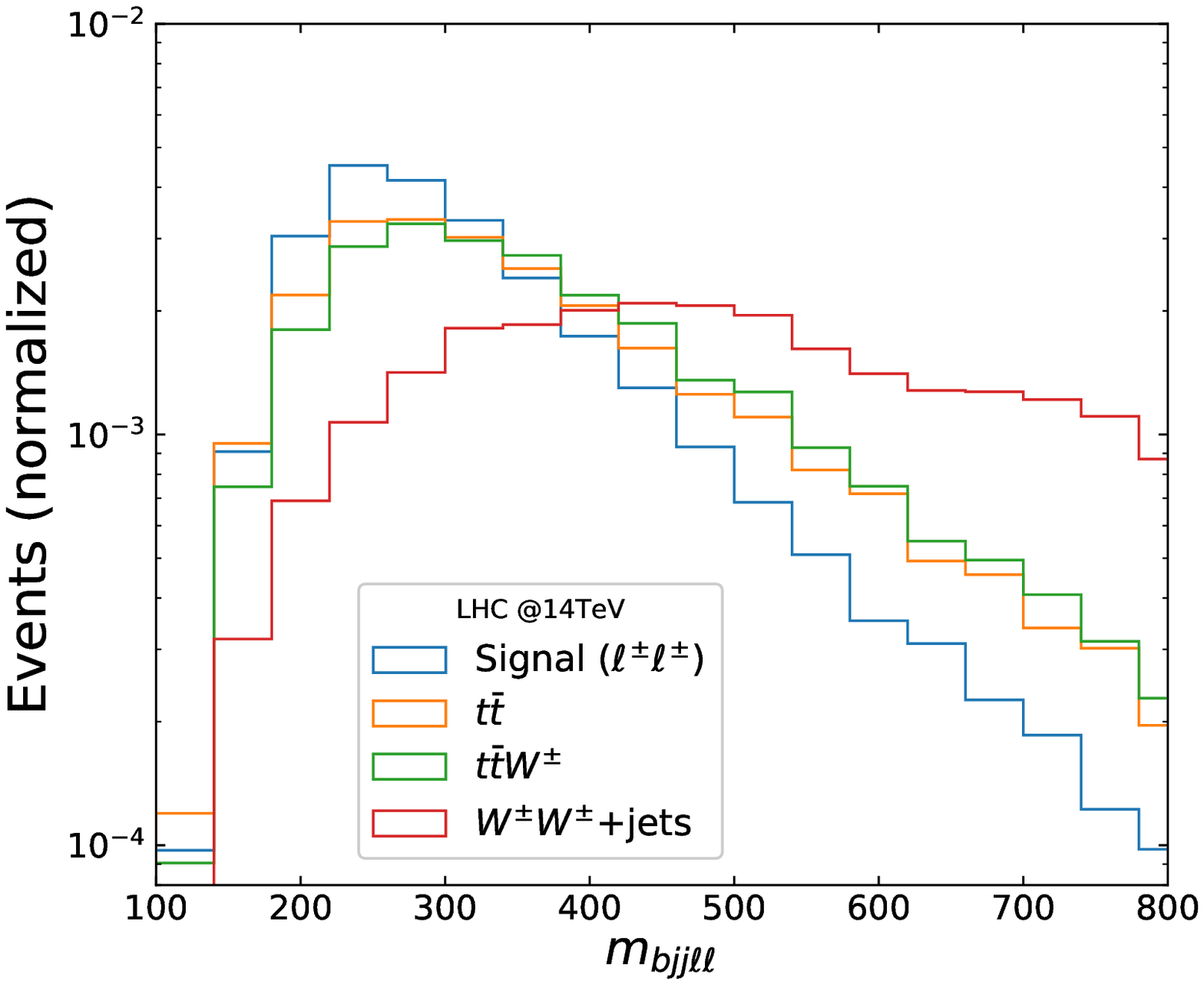}
  \caption{Kinematical distributions of signal $pp \to \ell^{\pm}\ell^{\pm}+2b+4j$ and backgrounds $t\bar{t}$, $t\bar{t}W^\pm$ and $W^\pm W^\pm$+jets at 14 TeV LHC. The benchmark point is $m_{N_4}=m_{N_5}=15$ GeV and $V_{\ell N}=V_{\mu N}=1$.}
  \label{distributions}
\end{figure}

In Fig.~\ref{distributions}, we present the kinematical distributions of the signal and SM background events at the 14 TeV LHC. We can see that the backgrounds $t\bar{t}$ and $t\bar{t}W^\pm$ have less events with same-sign muons than our signal (top-left panel). Since there are no neutrinos in our signal process, the missing transverse energy $\slashed E_T$ of the signal events are smaller than that of background events (top-right panel). Besides, we note that the two same-sign muons in the signal events come from the decay of the same top quark and hence tend to be closer. While two muons in the backgrounds arise from the decays of different parent particles.  Thus they separate from each other, which can be seen in the distribution of $\Delta R_{\ell\ell}$ (lower-left panel). Furthermore, the reconstruction of top quark also plays an important role in discriminating the signal from backgrounds. To do so, we present the cluster mass $m_{bjj\ell\ell}$, in which the leading $b$ jet and two soft jets are used. This is because the jets from the top five-body decay $t \to b \ell^{+}\ell^{+}jj$ are averagely softer than those in the top quark decay $t \to bjj$. We can see that most of the signal events distribute around $m_{bjj\ell\ell} \sim 200$ GeV, which can be used to suppress $W^\pm W^\pm$+jets background events.

According to the above discussions, we apply the following cuts to select the signal events in our analysis:
\begin{itemize}
\item{Cut-1:} We require a pair of same-sign leptons in the final states. Each of leptons should satisfy $p_{T} (\ell) > 10$ GeV and $|\eta| < 2.8$.
\item{Cut-2:} At least 6 jets with $p_{T} (j) > 15$ GeV and $|\eta| < 3.0$ in the final states are required.
\item{Cut-3:} The missing transverse energy is required to satisfy $\slashed E_{T} < 25$ GeV.
\item{Cut-4:} We also require two lepton separation to be $0.4<\Delta R_{\ell\ell}<2.5$, as well as the lepton-jet separation $\Delta R_{\ell j} > 0.4$ and jet-jet separation $\Delta R_{jj} > 0.4$.
\item{Cut-5:} We demand at least one $b$-jet with $p_T(b)>20$ GeV in the final states.
\item{Cut-6:} We require the reconstructed invariant mass $m_{bjj\ell\ell}$ lie in the range of $[130, 350]$ GeV.
\end{itemize}

\begin{table}
\centering
\begin{tabular}{l|c|c|c|c}
\hline\hline
 & $t\bar{t}$ & $t\bar{t}W^{\pm}$ & $WW+$jets & signal \\ \hline
Cut-1 & 8.83 & $2.90\times10^{-2}$ & $3.25\times10^{-3}$ & 22.7  \\ \hline
Cut-2 & 0.891 & $4.70\times10^{-3}$ & $1.88\times10^{-4}$ & 6.48  \\ \hline
Cut-3 & 0.128 & $3.95\times10^{-4}$ & $1.12\times10^{-5}$ & 2.95  \\ \hline
Cut-4 & $7.40\times10^{-2}$ & $2.39\times10^{-4}$ & $5.23\times10^{-6}$ & 2.35  \\ \hline
Cut-5 & $4.66\times10^{-2}$ & $1.66\times10^{-4}$ & $7.27\times10^{-7}$ & 1.97  \\ \hline
Cut-6 & $3.38\times10^{-2}$ & $1.10\times10^{-4}$ & $1.45\times10^{-7}$ & 1.35 \\
\hline\hline
\end{tabular}
\caption{Cutflow of cross sections of the signal process $pp \to \ell^{\pm}\ell^{\pm}+2b+4j$ and background processes $pp \to t\bar{t}, t\bar{t}W^\pm, WW$+jets at the 14 TeV LHC. The cross sections are in the unit of pb. The benchmark point is chosen as $m_{N}=15$ GeV and $V_{\ell N}=1$.}
\label{tab:cutflow14}
\end{table}

In Table~\ref{tab:cutflow14}, we show the cutflow of cross sections of the signal process $pp \to \ell^{\pm}\ell^{\pm}+2b+4j$ and background processes $pp \to t\bar{t}, t\bar{t}W^\pm, WW$+jets at the 14 TeV LHC. It can be seen that the $t\bar{t}$ process is the dominant background, which is followed by $t\bar{t}W^\pm$ process.
After imposing the requirement of same-sign muons, the cross section of $t\bar{t}$ process is reduced to the same order as that of signal process. Then, the small $\slashed E_T<25$ GeV, large jet multiplicity $N(j)\geq 6$ and $0.4<\Delta R<2.5$ conditions further suppress the cross sections of all backgrounds by an order of ${\cal O}(10^2)$. In the end, the cluster mass cut will remove the $WW$+jets background events and reduce the $t\bar{t}W^\pm$ background events to negligible level.
Thus with the above cuts, we expect we can have a promising sensitivity of probing our parameter space because of very few background events.


In order to estimate the signal significance ($\alpha$), we adopt the following formula,
\begin{align}
\alpha=S/\sqrt{B+(\beta B)^2}
\end{align}
in which $S$ and $B$ stand for number of signal and background events after our cuts, respectively. ${\cal L}$ is the integrated luminosity of the collider. It should be mentioned that the main systematic uncertainty is related to the misidentified-lepton backgrounds. By combining other sources, we include a systematic uncertainty of $\beta=5\%$ in our calculations.

\begin{figure}[th]
\centering
\includegraphics[width=8cm,height=7cm]{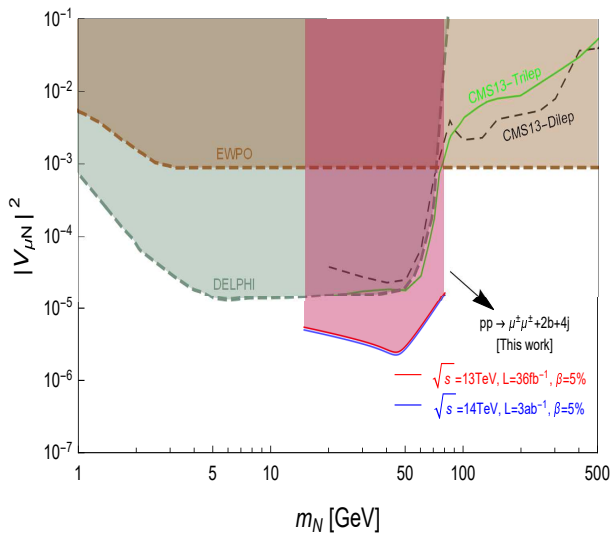}
\caption{The contour of $2\sigma$ exclusion limits from search of signal events $pp \to 2\mu^{\pm}+2b+4j$ on the plane of $m_N$ versus $|V_{\mu N}|^2$. Other limits are also shown: the electroweak precision measurements~\cite{delAguila:2008pw}, search for neutral heavy leptons produced in $Z$ decays at LEP~\cite{Abreu:1996pa}, and searches of same-sign leptons~\cite{Sirunyan:2018xiv} and trileptons events~\cite{Sirunyan:2018mtv} at the LHC.}
\label{fig:mumucontour}
\end{figure}
In Fig.~\ref{fig:mumucontour}, we present $2\sigma$ exclusion limits of the signal process $pp \to t\bar{t} \to \mu^{\pm}\mu^{\pm}+2b+4j$ on the plane of $m_N$ versus $|V_{\mu N}|^2$. To compare our results with others, we also plot the limits from electroweak precision measurements, LEP search for neutral heavy leptons produced in $Z$ decays, and LHC searches of same-sign leptons and trileptons events. It can be seen that the light-heavy neutrino mixing parameter $|V_{\mu N}|^2 >2.5\times 10^{-6}$ can be excluded at $2\sigma$ level in the masses of heavy neutrinos between 15 GeV and 80 GeV at 13 TeV LHC with the luminosity of 36 fb$^{-1}$, which is stronger than other existing bounds. Such a limit will be further improved to $|V_{\mu N}|^2 >2.3\times 10^{-6}$ at future HL-LHC with the luminosity of 3000 fb$^{-1}$.


\begin{figure}[th]
\centering
\includegraphics[width=8cm,height=7cm]{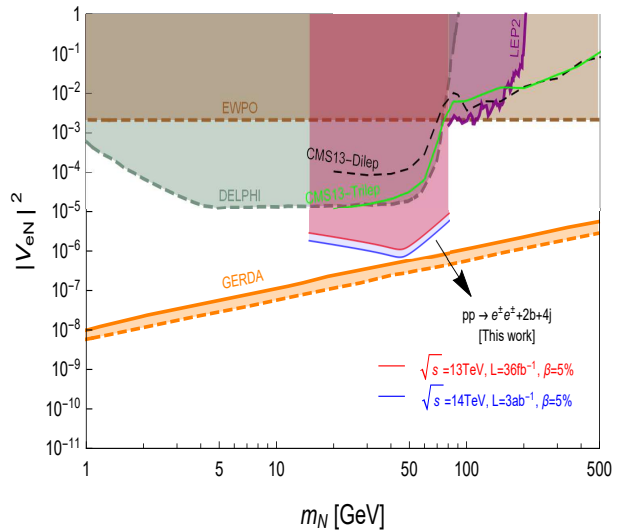}
\caption{Same as Fig.~\ref{fig:eecontour}, but for the signal process $pp \to 2e^\pm+2b+4j$. The indirect constraint from the search of neutrinoless double beta decay with the GERDA experiment~\cite{Agostini:2018tnm} is plotted as well.}
\label{fig:eecontour}
\end{figure}
Besides, we also studied the sensitivity of the electron channel $pp \to e^\pm e^\pm+2b+4j$ in Fig.~\ref{fig:eecontour}. The exclusion limits on the mixing parameter $|V_{e N}|^2$ is of the same order as $|V_{\mu N}|^2$, which is $|V_{e N}|^2 > 1.1\times 10^{-6}$ at 13 TeV LHC with the luminosity of 36 fb$^{-1}$ and $|V_{e N}|^2 > 7.2 \times 10^{-7}$ at the HL-LHC. This bound is weaker than that from the current GERDA search of neutrinoless double beta decay of $^{76}$Ge~\cite{Agostini:2018tnm}. However, it should be mentioned that $0\nu\beta\beta$ constraint is usually model dependent. For example, there is no such a limit in models with nearly-degenerate heavy Majorana neutrinos~\cite{Bray:2007ru}. Also, in radiative neutrino mass models, the cancelation effects between different amplitudes induced by the Majorana phase~\cite{Helo:2015fba} can significantly weaken this bound. Therefore, it is still necessary to analyze the electron channel while performing a complementary search for heavy neutrinos at colliders.

Finally, we comment that our analyses have not included pileup effects. In a fully realistic analysis it is important to include the effects from pileup and the effects of applying the appropriate pileup removal techniques~\cite{Cacciari:2007fd,Krohn:2013lba, Berta:2014eza}. For the analyses presented in this paper, however, events are selected with two hard same-sign leptons so we expect such additional considerations will not significantly alter the results.

\section{Conclusions}
Many low-scale Type-I seesaw models predict the heavy neutrinos below TeV scale. In order to test the mass generation mechanism and flavor structure of neutrinos in these models, it is crucial to measure the light-heavy neutrino mixing parameters $V_{i N}$ and masses of heavy neutrinos $M_N$. In this paper, we proposed to probe the heavy Majorana neutrino in top quark neutrinoless double beta decay through $t\bar{t}$ production process, which gives a distinctive signature $\ell^{\pm}\ell^{\pm}+2b+4j$ at the LHC. After performing the detector level simulation, we find that the mixing parameters $|V_{e N}|^{2}>1.1 \times 10^{-6}$ and $|V_{\mu N}|^{2}> 2.5 \times 10^{-6}$ in the mass range of 15 GeV$<m_N<$ 80 GeV can be excluded at $2\sigma$ level at 13 TeV LHC with the luminosity of 36 fb$^{-1}$, which have surpassed other existing collider bounds. The future HL-LHC will further improve these limits. Therefore, we conclude that searching for the neutrinoless double beta decay of top quark will provide a new better way to probe the heavy Majorana neutrino in the mass range of 15 GeV$<m_N<$ 80 GeV at the LHC.

\section{acknowledgments}
This work is supported by the National Natural Science Foundation of China (NNSFC) under grant Nos. 11847208, 11875179, 11805161, 1705093, as well as Jiangsu Specially Appointed Professor Program.

\end{document}